\documentclass[10pt]{article}
\usepackage{epsfig}

\newcommand{\be}{\begin{equation}}
\newcommand{\ee}{\end{equation}}
\newcommand{\ba}{\begin{eqnarray}}
\newcommand{\ea}{\end{eqnarray}}

\begin{document}

\title{L\'evy Stable Law Description of the Intermittent Behavior 
in Pb+Pb Collisions at 158 AGeV/c}

\author{
A.M. Tawfik and E. Ganssauge \\ \\
{\small FB Physik, Marburg University, Renthof 5,
D-35032 Marburg, Germany}}

\date{}
\maketitle
\parindent0mm

\begin{abstract}
The factorial moments (FM) of multiplicity distributions
in one- and two-dimensions are studied for Pb+Pb collisions at
158 AGeV/c. The results are compared with FRITIOF, RQMD and VENUS
predictions. In the relation between FM and the number of
partitions, $M$, obvious {\it positive} intermittence exponents,
$\phi_q$, are observed. The exponents, $\phi_q$, are compared with
the anomalous dimensions, $d_q$. Then they are expressed in terms
of L\'evy stable indices, $\mu$. Meanwhile the experimental data
within the interval, $2<\eta <6$, fulfills the requirements of the
L\'evy stable region ($0 \le \mu \le 2$), the results from the
other intervals are away from this region. All their
corresponding indices have negative values. \\

\end{abstract}

\section{Introduction}
\label{sec:1}
Since the observations of JACEE collaboration \cite{jacee} and
the pioneer works of Bai{\l}as and Peschanski \cite{Bia86,Bia88}
around mid of 1980's, the intermittent behavior in multi-particle
production has increasingly gained full credibility among the
high-energy physicists. The concept {``intermittence"} itself is
primarily used to describe the fluctuations observed in the
particle density distributions. In many experiments, the {\it
large dynamic} fluctuations at the final state of particle
production has been confirmed. To get a satisfactory explanation
for the power-law behavior of FM with the partition number, $M$,
many theoretical approaches are proposed. For example,
self-similar branching, QCD parton cascade, multi-particle
cascade, second-order phase-transition, etc. Also the effects of
Bose--Einstein correlations on the two-particle correlation
functions \cite{Cur18,Cap18,Biy20,Neu21} may explain this
behavior. The randomization and the hadronic \'Cerenkov radiation
are suggested to be partially responsible for such power-scaling
behavior. Also, if QGP has been produced, the effects of the soft
and hard collisions, the clustering and resonance decay and the
showering in the QGP-surrounding manifest themselves in forms of:
multiplicity fluctuation, power-scaling behavior, self-similar
branching, jet structure, etc. Some of these effects are
responsible for a kind of {\it canonical} fluctuations in the
particle production.

In this paper, the behavior of FM is studied for Pb+Pb collisions
at 158~AGeV/c with successive partitions in one- (pseudo-rapidity
$\eta$ or azimuthal angles $\phi\in\{0,2\pi\}$), and
two-dimensions ($\eta$ and $\phi$). The produced particles are
classified into four groups according to their pseudo-rapidity
values ($3<\eta<4$, $3<\eta<5$, $2<\eta<5$ and $2<\eta<6$). The
observed intermittence exponents, $\phi_q$, are compared with the
anomalous dimensions, $d_q$ (R\'enyi dimensions, ${\cal R}_q$).
After that, they are plotted in the L\'evy space. \\

This article is organized as the following: The experimental
material and the simulation by using the three events generators,
FRITIOF, RQMD and VENUS are complemented in Section \ref{sec:2}.
The definitions and formalism are given in Section \ref{sec:3}.
The results of FM and the intermittence exponents, $\phi_q$, in
one- and in two-dimensions are introduced in Sections \ref{sec:4}
and \ref{sec:5}, respectively. The relations between the
intermittence and the anomalous exponents and their expressions
in terms of L\'evy indices, $\mu$, are discussed in Section
\ref{sec:levy}. Section \ref{sec:7} contains some remarks and the
final conclusions.

\section{Experimental Material}\label{sec:2}

The data used in this paper is retrieved from some of the
Pb-chambers irradiated at CERN-SPS during 1996 for the EMU01
collaboration. This material was consummately measured and
analyzed at the Marburg University by using our measuring system
MIRACLE Lab \cite{TawDs,Taw241,Taw242}. The collisions have been
recorded with $^{82}$Pb-beam accelerated to incident momentum of
158 AGeV/c and then fired towards stationary lead foil with
thickness $\sim$ 250 $\mu$m. The lead foil was positioned in the
front of seven consecutively arranged plastic sheets coated on
both sides by the nuclear emulsion, FUJI ET-7B. Because of the
relatively high transverse momentum, most of, if not all, produced
particles and fragmentations are emitted within a narrow forward
cone. Therefore, their tracks can be registered within the
forward emulsion sheets. The exposure process is mainly
controlled by counting the heavily ionizing particles by using
scintillator and discriminator with relatively high threshold
settings. The number of beam-particles is registered by using an
additional counter installed behind the emulsion chamber.  Scale
and driver electronics (CAEN N145) are used to build up pulses to
be transferred to the SPS-{``kicker''} magnet. The kicker magnet
consequently removes the beam, if the particle number transcends
3000 per spot. The beam density is $\sim 5\cdot 10^2$
nuclei/cm$^2$.

The scanning efficiency in the emulsion chambers is $0.75 \pm
0.05$ \cite{EMU01-1}. The emulsion sensitivity for singly charged
particles is as good as 30 grains per 100 $\mu$m. Depending on
the incident momentum, the polar angle can be calculated. For the
actual data, we get polar angle of 1.3 {mrad}. Depending on
topology of the microscope's field of view, particles with space
angles, $\theta< 30^{\circ}$ ($\eta = -\ln\tan(\theta/2)> 1.32$)
can be acquired. Obviously, most of produced particles are singly
charged. It is expected that they are mixed with contamination of
$e^- - e^+$ pairs from Dalitz decay and $\gamma$-conversion. The
possible {\it overestimation} of the observed particles density
has been determined as $\sim 2\%$ \cite{TawDs}. As a reason of
the reconstruction algorithm \cite{Taw242} applied for the
measuring system MIRACLE Lab, the tracks of these electrons are
{\it completely} disregarded. In comparison with {\it human}
operators, it is found that MIRACLE Lab has an efficiency up to
$96\%$ \cite{Taw242}. Large part of this $4\%$ discrepancy is
obviously coming from the particles emitted with large space
angles ($\theta \ge 30^\circ$). The missing measurements,
frequent scattering, unresolved close-pairs and pair production
represent additional sources for this discrepancy \cite{Taw241}.
Therefore, we could consciously suggest to renounce the
discussion of the effects of $\gamma$-conversion on FM ({see}
Section \ref{sec:gamma1d} and Section \ref{sec:gamma2d} for more
details).

\subsection{Simulation of the experimental data}
\label{sec:21}

The investigation of FM is performed for the collisions with
restricted particle multiplicity\footnote[1]{
This is a restriction on the multiplicity ({see} Section
\ref{sec:3} for the effects of the multiplicity restriction on FM \cite{WU92}
} 
$\ge 1200$. From these
events, only the particles emitted within predefined
pseudo-rapidity intervals ($3<\eta<4$, $3<\eta<5$, $2<\eta<5$ and
$2<\eta<6$) are taken into account.  These $\eta$-intervals are
chosen at and around the mid-rapidity region. Due to the chiral
symmetry breaking, the produced particles are mainly pions,
$\pi$'s. Therefore, the restriction on the $\eta$-values is
necessary to study the phase-transition, which, in turn, should
be restricted to the produced particles only. Therefore, the
narrow $\eta$-intervals are the suitable ones to study the
produced particles and consequently the phase-transition. Once
again, here we used another restriction on the considered
phase-space.\footnote[2]{
The influences of the restriction on the considered
    phase-space to get {\it flat} distributions of FM are discussed in
    Section \ref{sec:3}.
}
On the azimuthal angles, there is no restriction
at all, (for each $\Delta\eta$ group, the considered
$\phi$-window is allowed to take any real value within the
available spectrum, $\{0,2\pi\}$).

For the event generators FRITIOF 7.02 \cite{fritiof1}, RQMD 2.1
\cite{rqmd1} and VENUS 4.12 \cite{venus1}, the input parameters
are adjusted for the destination to produce simulated events with
total multiplicity not smaller than $1200$. Each simulated event
has to go through different cutting processes, in order to
simulate, as good as possible, the {``real''} events (Section
\ref{sec:2}). After these operations, the number of particles
observed within the space angle, $\theta < 30^{\circ}$, has to be
$\ge 1200$. The criteria used for the data cutting, can be taken
from \cite{TawDs,Taw241,Taw242}:
\begin{enumerate}
\item $\beta =\vec{p}/E > 0.7$. In practice, we can mainly measure
      singly charged particles. More than $90\%$ of them are charged $\pi$'s.
      The number of highly charged particles is relatively small \cite{EMU01-1}.
\item $\eta = -\ln\; \tan(\theta/2)> 1.32$. The space angle is restricted within
      $\theta < 30^{\circ}$ depending on the geometry of the emulsion chamber,
      and of the microscope's field of view.
\item Charged particles.  The emulsion is sensitive for
      the charged particles only.
\end{enumerate}

\section{Definitions and Formalism}
\label{sec:3}

If the phase-space, $\Delta\eta$, is split into $M$ bins of equal
sizes, $\delta\eta=\Delta\eta/M$, the {\it exclusive} scaled
factorial moments \cite{Bia86,Bia88} are given as

\be F_q(M)^{\mbox{excl}} = M^{q-1} \sum\limits_{m=1}^{M}
       \frac{\left<n_m(n_m-1)\cdots(n_m-q+1)\right>}
            {\bar{n}(\bar{n}-1)\cdots(\bar{n}-q+1)},
\label{e:1}
\ee

$n_m$ is the multiplicity in $m$-th bin and $\bar{n}$ is the
average multiplicity in the whole $\Delta\eta$-window. With
successive partition, $\Delta\eta \rightarrow 0$, a power-law
scaling can describe the fluctuations in the particle production,

\be F_q(M) = M^{q-1} \sum\limits_{m=1}^{M}
\frac{\prod\limits_{i=1}^{q}\;\;\int\limits_{\eta_{m-1}}^{\eta_m}
      d \eta_i \; \rho_n^{(q)}(\eta_1,\eta_2,\cdots,\eta_q)  }
     {\prod\limits_{i=1}^{q}\;\;\int\limits_{\eta_{0}}^{\eta_0+\Delta\eta}
      d \eta_i \; \rho_n^{(q)}(\eta_1,\eta_2,\cdots,\eta_q) },
\label{e:2}
\ee

$\rho_n^{(q)}(\cdots)$ is the pseudo-rapidity density in a
sub-interval with $q$ particles investigated in an event with $n$
total multiplicity. $\eta_0$ is the minimum pseudo-rapidity in
$m$-th bin and $\eta_m$ is the pseudo-rapidity value in this bin
($\eta_m=\eta_0+m\cdot\Delta\eta/M$). Then the factorial moments
in $m$-th bin are to be given as

\be F_q(M)=
\frac{\prod\limits_{i=1}^{q}\;\;\int\limits_{\eta_{m-1}}^{\eta_m}
      d \eta_i \; \rho_n^{(q)}(\eta_1,\eta_2,\cdots,\eta_q) }
     {\prod\limits_{i=1}^{q}\;\;\int\limits_{\eta_{0}}^{\eta_0+\Delta\eta}
      d \eta_i \;\rho_n^{(q)}(\eta_1,\eta_2,\cdots,\eta_q) }.
\label{e:3} \ee

In the case that there is no correlation between the produced
particles, FM, in the whole $\Delta\eta$-interval, can be given
as

\be F_q(M)  =  M^{q-1} \sum\limits_{m=1}^{M}
               \left(\;\;\int\limits_{\eta_{m-1}}^{\eta_m}
               d \eta_m\;\;\bar{P}_n(\eta) \right)^q ,
\label{e:4} \ee

$\bar{P}_n(\eta)  = \rho_n(\eta)/\bar{n}$ is the normalized
distribution of pseudo-rapidity density in a bin of with size
$\eta$. When $\eta\rightarrow \Delta\eta$, we get the maximum
value

\be F_q(M)^{\mbox{max}} =
\Delta\eta^{q-1}\int\limits_{\eta_0}^{\eta_0+\Delta\eta_m} d \eta
\;\; \left[\bar{P}_N(\eta)\right]^q. \label{e:5} \ee

When $F_q(M)^{\mbox{max}}=1$, the single-particle distribution in
the $\Delta\eta$-window will be a plane one\footnote[3]{
A powerful method to produce {\it flat} particle density
distributions from {\it non-flat} ones can be taken from \cite{Hove89}.}
 (a distribution with non-statistical fluctuations).

In this section, we have discussed the {\it exclusive} FM and
their limitations to describe the multiplicity fluctuations. In
the next section, we pass to their possible applications to
describe the particle production in heavy-ion collisions.

\subsection{Factorial moments and intermittence exponents}
\label{sec:31}

For many reasons\footnote[4]{
Mainly, there are two reasons: the limitation on the
    considered pseudo-rapidity, in order to get flat particle density
    distributions and the limitation on multiplicity \cite{EHS93}, in
    order to avoid any possible {\it dynamical} correlation between
    the produced particles.},
the direct calculation of the {\it
exclusive} FM is not easy, especially for high statistics.
Therefore, it has been suggested \cite{Bia86,Bia88,Peschn90} to
implement another kind of FM, namely the {\it inclusive} FM,

\be F_q(M)^{\mbox{incl}}= M^{q-1} \sum\limits_{m=1}^{M}
       \frac{\left<n_m(n_m-1)\cdots(n_m-q+1)\right>}{\bar{n}^q}.
\label{e:6} \ee

The value, $\bar{n}$, is coming out by dividing the observed
multiplicity by the bin number, $M$, and then by the number of
events, $N$. The exclusive and inclusive FM are the same, if the
particle production can be described by Poisson function. In
principle, the particle density distribution depends on the
dynamics of the interacting system (mass, energy, number of
events, whether there is any restriction in the data sample,$^e$
etc.).

\be F_q^{\mbox{incl}} = \frac{\bar{n}(\bar{n}-1) \cdots
   (\bar{n}-q+1)}{\bar{n}^q}\cdot F_q^{\mbox{excl}}.
\label{e:7}
\ee

Then in the real case, the difference between both types of FM is
actually the long-range correlations, which appear as a reason of
the strong effects of the multiplicity on the correlation
functions \cite{WU92,ChWa1,ChWa2}.

By definition \cite{Bia86,Bia88,Peschn90,Peschn91}, FM used in
the particle production are

\be <F_q(M)> = \left<\frac{<n_k(n_k-1) \cdots
                 (n_k-q+1)>}{<n_k>^q}\right>,
\label{e:8} \ee

$<n_k>$ is the average multiplicity in $k$-th bin of size
$\delta\eta$. According to the {\it self-similar density
fluctuations}, the successive partitions of the pseudo-rapidity,
$\Delta\eta$, leads to the following power-law dependency

\be <F_q(M)> \;\; \propto \;\; M^{\phi_q} . \label{e:9} \ee

The exponents, $\phi_q$, are called the {``intermittence
exponents''}, which practically can be retrieved from the slopes
of the relations between $\log F_q$ and $\log M$. They are also
related to the {``anomalous dimensions''} \cite{Lipa89}

\be d_q = \frac{\phi_q}{q-1}. \label{e:10} \ee

The fractal R\'enyi dimensions \cite{Busc88} are defined as
functions of the {``anomalous dimensions''} $d_q$

\be {\cal R}_q = {\cal R} \cdot (1-d_q), \label{e:11} \ee

The constant, ${\cal R}$, is the topological R\'enyi dimension.
For {\it multi-fractal} processes, $d_q$ can be given as
functions of the orders, $q$, only (Section \ref{sec:61} and
Section \ref{sec:62}). For {\it mono-fractal} processes (Section
\ref{sec:levy}), $d_q$ are constant. Actually, R\'enyi dimensions
can be used to measure the randomization in the produced
particles. For the total randomization, $\phi_q=0$, $d_q=0$, and
consequently ${\cal R}_q={\cal R}$.

\section{Results in One-Dimension}
\label{sec:4}

\begin{figure}[htb]
\centerline{\epsfxsize=6cm \epsfbox{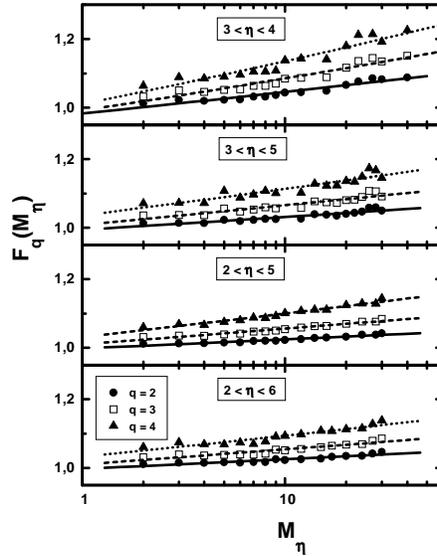}} \vspace*{-12pt} 
\caption[]{\small \it The factorial moments, $F_q$, for
  the orders, $q=\{2,3,4\}$, are plotted as functions of {\it small}
  partition number, $M_\eta$, in $\eta$-dimension. The data used here
  are Pb+Pb collisions at 158 AGeV/c with total multiplicity
  $\ge 1200$.
  The different $\Delta\eta$-windows are applied above the whole
  region of pseudo-rapidity ($\eta > 1.32$). Only the secondary
  particles emitted within these windows are taken into consider. The
  data are fitted as straight lines. Their slopes are the
  {``intermittence exponents''}~$\phi_q$.
\label{fig1} }
\vspace*{20pt}
\end{figure}

\begin{figure}[htb]
\centerline{\epsfxsize=6cm \epsfbox{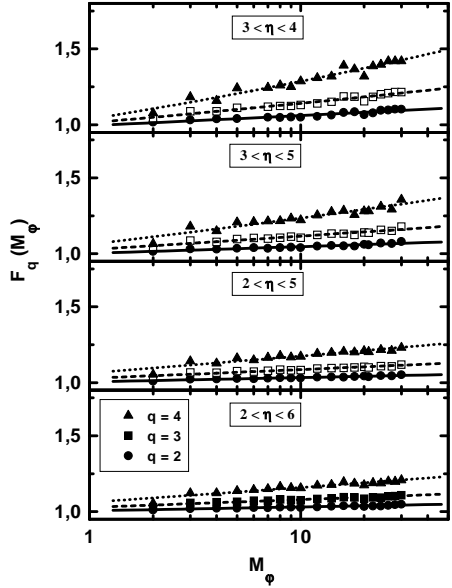}} \vspace*{-12pt} 
\caption[]{\small \it The
factorial moments, $F_q$, for the orders, $q=\{2,3,4\}$, are
plotted as functions of {\it small} partition number, $M_\phi$, in
$\phi$-dimension. As in {Fig.~\ref{fig1}} the data are fitted as
straight lines. Correspondingly their slopes give the component
of {``intermittence exponents''} in $\phi$-dimension, ~$\phi_q$.
\label{fig2}}
\vspace*{20pt}
\end{figure}

\begin{figure}[htb]
\centerline{\epsfxsize=6cm \epsfbox{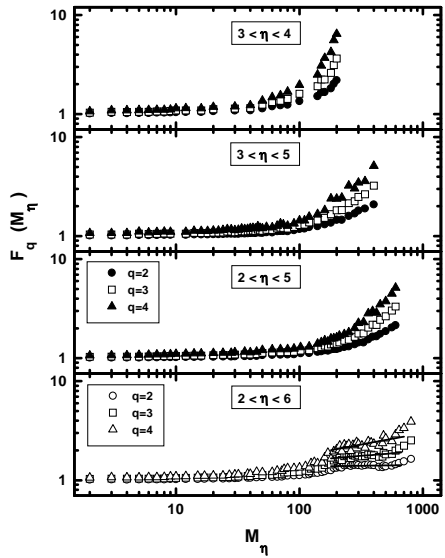}} \vspace*{-12pt} 
\caption[]{\small \it The
factorial moments, $F_q$, for the orders, $q=\{2,3,4\}$, are
depicted as functions of {\it large} partition number, $M_{\eta}$,
in $\eta$-dimension. The partition process is repeated as long as
the FM-values increase. An exponential growth of FM is noticed
for relatively large $M_{\eta}$. Within the largest
$\Delta\eta$-interval, we notice that the increasing of FM is
saturated.  \label{fig3}}
\vspace*{20pt}
\end{figure}

\begin{figure}[htb]
\centerline{\epsfxsize=6cm \epsfbox{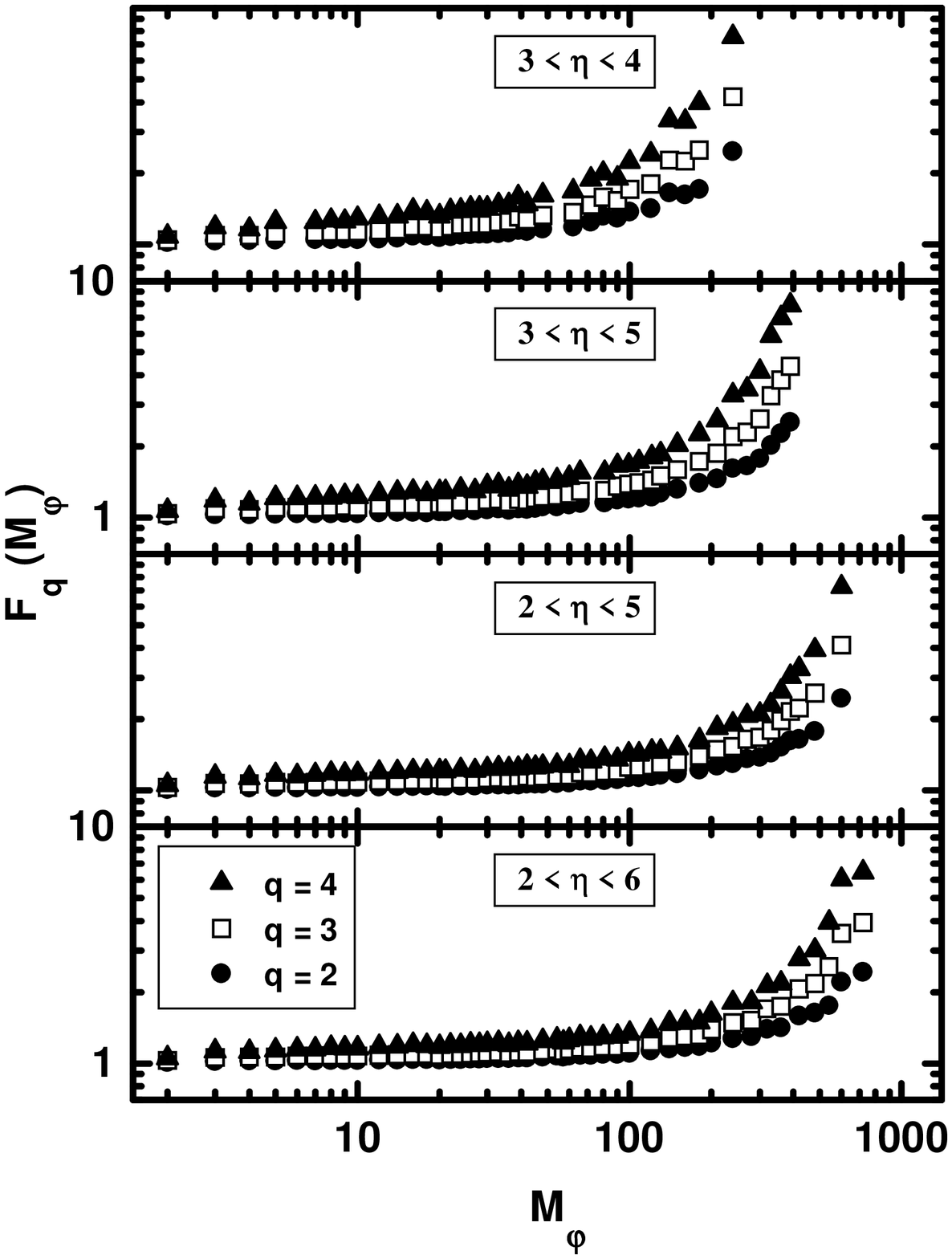}} \vspace*{-12pt} 
\caption[]{\small \it The
factorial moments, $F_q$, for the orders, $q=\{2,3,4\}$, as
functions of relatively {\it large} partition number, $M_{\phi}$,
in $\phi$-dimension. As in {Fig.~\ref{fig3}} the partition process
is repeated as long as the FM-values increase. Here there is no
evidence for the saturation region.  \label{fig4}}
\vspace*{20pt}
\end{figure}

In both {``real''} and {``simulated''} collisions,
pseudo-rapidity intervals are successively split into $M$ {\it
equal} sub-intervals (bins). In each such bins, the multiplicity
and then the corresponding $q$-order FM are calculated according
to {Eq.\ (\ref{e:8})}. {Fig.~\ref{fig1}} shows $F_q$ as functions
of $M_{\eta}$ for the orders, $q=\{2,3,4\}$, and for the different
$\Delta\eta$-intervals given in Section \ref{sec:21}. The
underscore in $M_{\eta}$ means that the partition has been
performed in $\eta$-dimension. The splitting in the other
dimension is completely avoided here. We notice that all pictures
show obvious {\it positive} intermittence slopes, $\phi_q$. Also,
it is to realize that the FM-values increase with the decreasing
of $\Delta\eta$-intervals. Another remarkable finding noticed here
is that the slopes increase with the increasing of the orders, $q$
({review} {Eq.\ (\ref{e:8})}). The analysis depicted in
{Fig.~\ref{fig2}} is for the partition in $\phi$-dimension.
Nearly, the same qualitative features as in previous figure are
noticed here. Through increasing the partition number, $M_{\eta}$
or $M_{\phi}$, the values of the corresponding FM raise fast, as
shown in {Fig.~\ref{fig3}} and {Fig.~\ref{fig4}}. The reasons for
this upwards increasing will be discussed later (Section
\ref{sec:41}). For now, the discussion is restricted only on the
phenomenological behavior of the increasing of FM with
$M_{\{\eta|\phi\}}$. In {Fig.~\ref{fig3}}, the smallest
$\delta\eta$ in all pictures is $\sim 0.01$, which seems to be
coincident with the detector resolution. Beyond this value, FM
reduce suddenly (not shown here), i.e.\ the upwards increasing of
FM (shown here) gets inverted. This boundary limit can be taken
as the low-value of the detector resolution. Until such point, we
can effectively count the produced particles and then correctly
calculate their FM. Beyond it, it is impossible to assure whether
the measured particles are {\it real} ones and consequently their
calculated FM must be associated with large statistical errors.
The range of the linear dependence in {Fig.~\ref{fig3}} becomes
larger with wider $\Delta\eta$-windows. Some kind of plateau begin
to appear with the increasing of $\Delta\eta$. Figure \ref{fig4}
depicts the same investigations as {Fig.\ \ref{fig3}}, but for
the partition in the $\phi$-dimension. Nearly, same qualitative
features as in {Fig.~\ref{fig3}} are noticed here. The only
difference is that there is no plateau in any curvature region. \\

\begin{figure}[htb]
\begin{center}
\begin{tabular}{cc}
\hspace{0.3cm}  FRITIOF &
\hspace{0.3cm}  RQMD\vspace{6pt}\\
\epsfysize=8cm\epsfbox{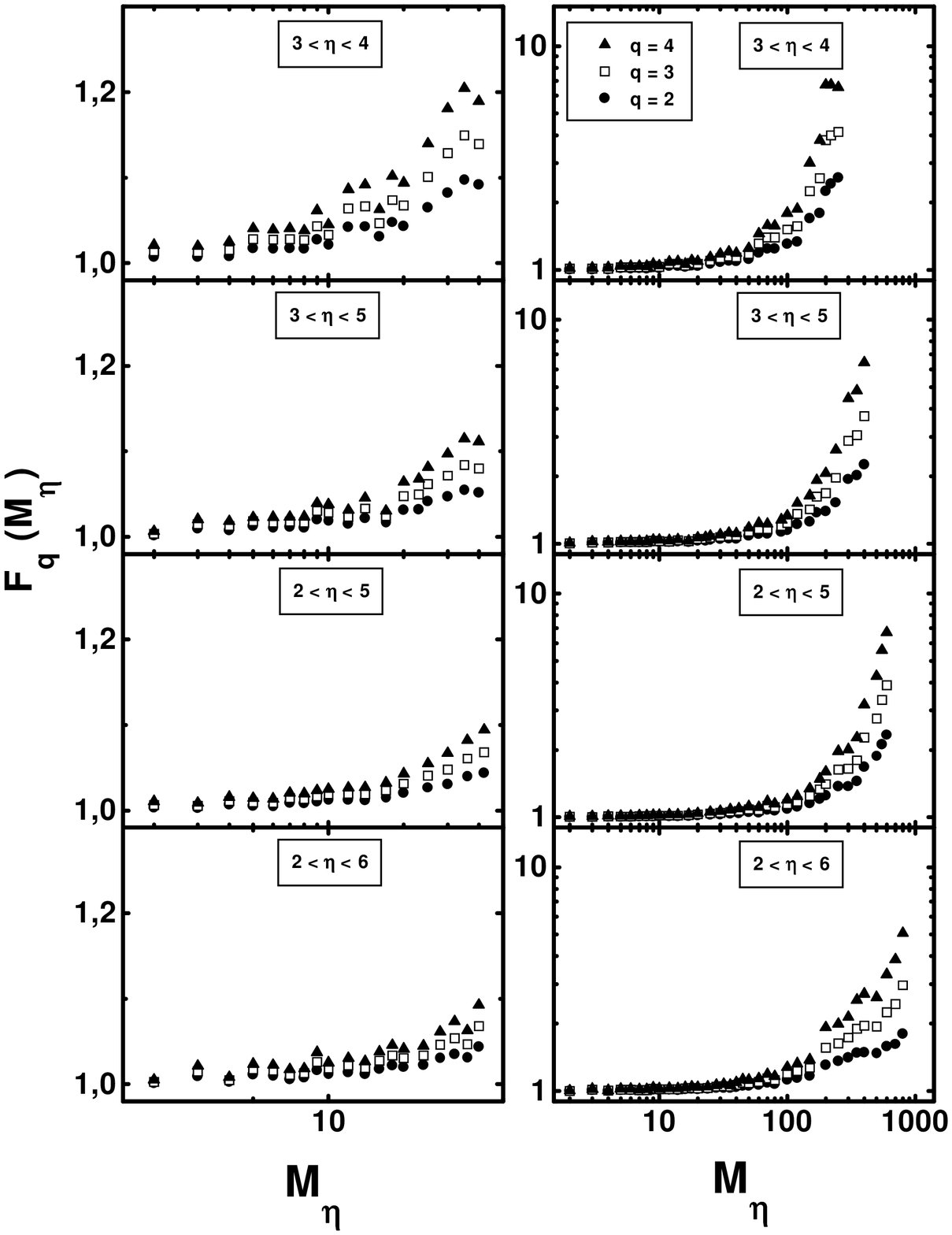}\quad
&
\quad\epsfysize=8cm\epsffile{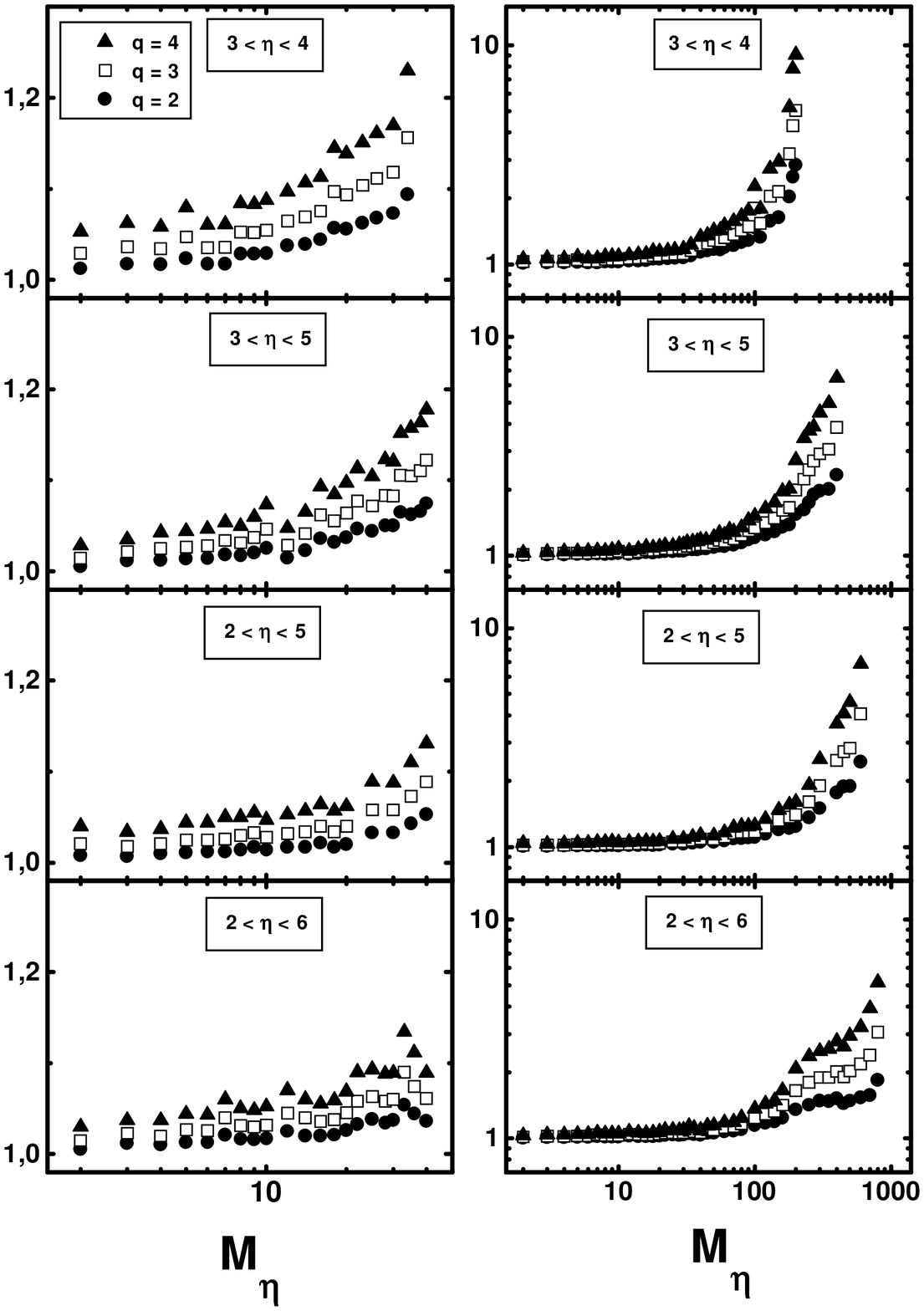} \\[0.25ex]
\end{tabular}
\vspace{-12pt}
\end{center}
\vspace*{6pt}
\begin{minipage}[t]{6.0cm}
\hspace*{88pt}VENUS\vspace{6pt}\\
\hspace*{24pt}{\epsfysize=8cm\epsfbox{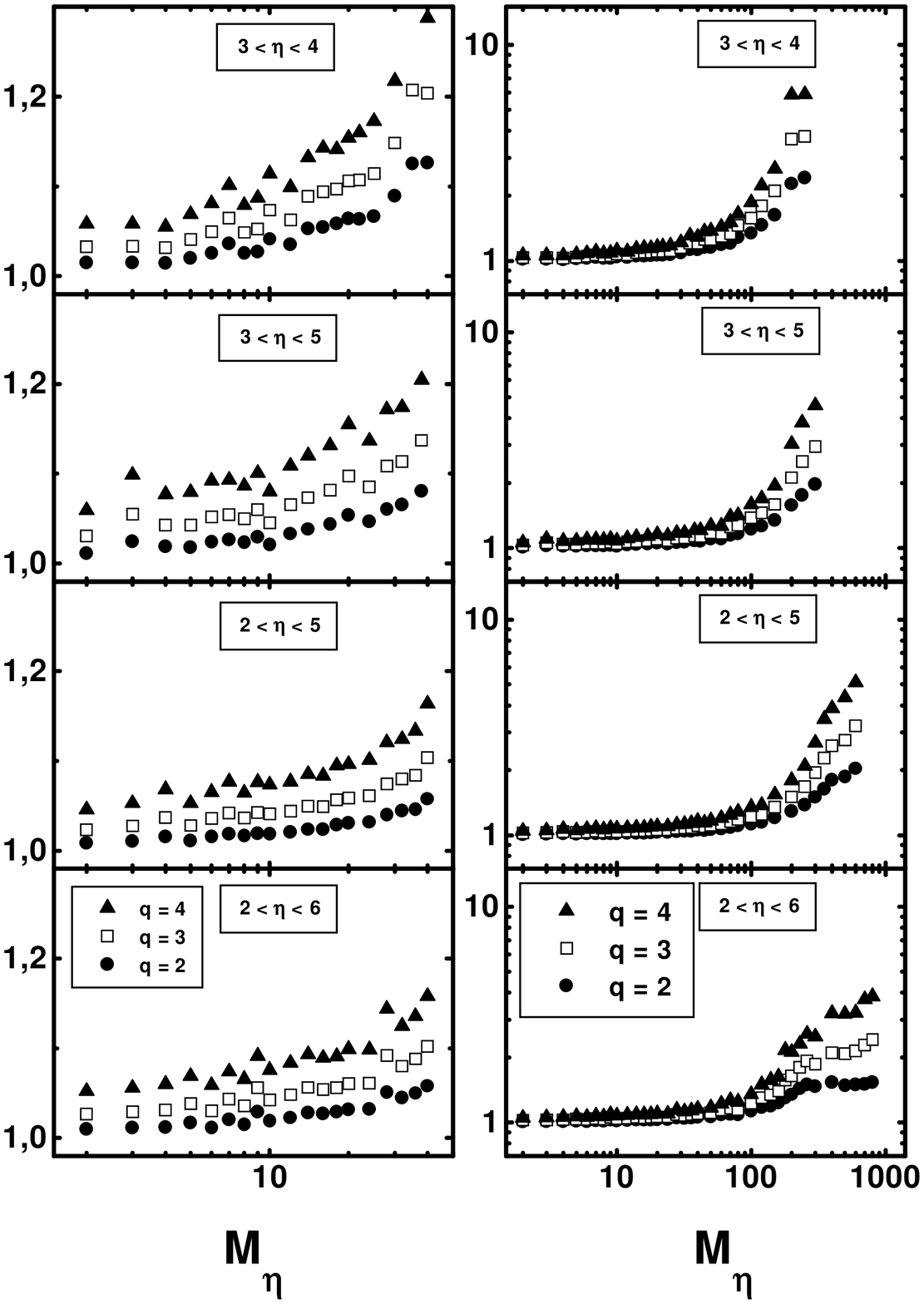}}
\end{minipage} \hfill
\vspace*{-76mm} \hspace*{75mm} \hfill
\begin{minipage}[t]{5.0cm}
\vspace*{2pt} \caption[]{\small \it The FM from FRITIOF, RQMD and VENUS
simulations depicted as functions of $M_{\eta}$. The left column
in each picture depicts the partition for relatively {\it small}
partition number, $M_{\eta}$. Obviously the experimental points
cannot be fitted as straight lines. The right column gives FM for
{\it large} $M_{\eta}$. In all pictures an exponential increasing
is noticed. In RQMD and VENUS simulations the increase of FM is
saturated especially in the largest $\Delta\eta$-interval. But
the results FRITIOF show no evidence for any saturation region.\label{Fig567}}
\end{minipage}
\vspace*{40pt}
\end{figure}

In addition to the large statistical errors, there are other
reasons for the negative dependence of $F_q(M_{\{\eta|\phi\}})$
on large $M_{\{\eta|\phi\}}$: The properties of FM themselves.
The FM used here can be applied, only if the following conditions
are simultaneously fulfilled:
\begin{itemize}
\item The order $q$ has an integer positive value.
\item The multiplicity in each bin ($n_m$) is larger
than or at least equal to $q$.
\item $q$ is greater than or at least equal to 2.
\end{itemize}
In {Fig.~\ref{fig3}} and {Fig.~\ref{fig4}}, we could expect that
the multiplicity, $n_m$, can be smaller than $q$. If the limit,
\hbox{$q < n_m$}, has been reached, we cannot correctly implement
{Eq.\ (\ref{e:8})}. The discussion of FM for large
$M_{\{\eta|\phi\}}$ should not, in the one hand, lead to the
conclusion that we would like to go through any argumentation of
the detector resolution. But, as we think, it is a {\it
worthwhile} finding to realize that our measuring system MIRACLE
Lab \cite{TawDs,Taw241,Taw242} is effectively able to determine
the emulsion resolution. This efficiency has been estimated (for
example, by testing its reproduction of the measured tracks, when
the emulsion plates are randomly rotated). In the other hand,
this discussion enabled us to refer to the limitations of the
used FM. \\

The predictions from the three events generators are illustrated
in {Fig.~\ref{Fig567}}. For FRITIOF simulation one notice that FM
have non-linear dependencies for relatively small $M_{\eta}$.
Also, the values of FM are not coincident with the experimental
ones shown in {Fig.~\ref{fig1}}. For large $M_{\eta}$, the
upwards increasing is to be observed until bin size
$\delta\eta\rightarrow 0.005$. Within these small bins, large
numerical uncertainties in the computations are obviously
expected. Therefore, these regions should be charily studied and
no serious conclusions according to their features should be
drawn. The saturation plateau here is too weak. Obviously, they
begin later and have smaller width than the experimental ones
({Fig.~\ref{fig2}}). From the results of RQMD calculations nearly
the same features as in previous picture are observed. The width
of the saturation plateau is larger than that in FRITIOF.
Although the saturation regions of the VENUS calculations seem to
be comparable with the corresponding regions in
{Fig.~\ref{fig2}}, they are still unable to completely predict the
experimental results. Obviously, the different discrepancies
between the experimental data and the three models become larger
with increasing $M_{\eta}$ ({see } {Fig.~\ref{fig14}}, later).
This may be explained due to the effects of Bose--Einstein
correlations, which are not included in any model. In addition to
this, during the experimental observations \cite{TawP12}, these
effects have not been considered. Another source for the
discrepancies may be the physics of the phase-transition which
has not been taken into account in any model.

\subsection{Upwards increasing}
\label{sec:41}

Going back to the gigantic upwards enhancement of FM with large
$M_{\{\eta|\phi\}}$. We would like to begin here with two points:
First, the results published in \cite{emu011}, in which the
authors claimed that there is no more increasing in FM in the
regions beyond the limit $\delta\eta < 0.1$. In this present
work, we clearly notice that FM still have an obvious linear
increasing, even up to the partition width $\delta\eta \approx
0.02$ ({see} {Fig.~\ref{fig1}} and {Fig. ~\ref{fig2}}). Beyond
this {\it narrow} bin size, the increasing of FM does not only
get ahead, but it becomes exponential. Second, a reasonable
explanation of the tremendous growth of FM for large
$M_{\{\eta|\phi\}}$ may be based on the {\it short-range}
conventional correlations \cite{Car89}, on the hadronic resonance
and cluster decay \cite{Cur18,Cap18}, on the Bose--Einstein
correlation, etc. In the following is another reason for this
exponential increase \cite{emu012,Yuan93}: Using trivial
geometric model, each projectile nucleon interacts along the
cylindrical tube in the target nucleus with more than one
nucleon. Therefore, for each of these sub-nucleus interactions,
there is a pseudo-rapidity distribution \cite{emu012} with
certain mid-rapidity depending on the typical characters of such
interacting systems. It depends also on the corresponding cross
section, which in turn depends on whether the {\it knock} takes
place between nucleons in face-to-face or between {\it eclipsed}
ones \cite{TawP13}. In addition to these sub-interactions, the
secondary particles (first and/or higher generations) produced
within the target material cause further interactions with each
other and with the target and projectile materials. These
secondary interactions produce additional pseudo-rapidity
distributions with their own mid-rapidity centers. Therefore, for
each of these {\it elementary} collisions, a {\it sub}
pseudo-rapidity center can be achieved. The superposition of
these elementary pseudo-rapidity values taken place in {\it
sub}-intervals $\delta\eta$, in which the original intervals
$\Delta\eta$ has been partitioned, makes the total {\it
effective} partition number, $M^{\rm eff}$, larger than $M$
($M^{\rm eff} \gg M$) \cite{emu012}. The reasons for these
additional effects can summarize as the following:
\begin{enumerate}
\item The successive elementary collisions cannot completely be coincident
with each others. Therefore the final summation of {\it
sub} pseudo-rapidity centers ($\sum\delta\eta$) produces one
distribution with many centers ($M^{\rm eff} \gg M$).
\item Even in the {\it ideal} case: The superposition produces only one
distribution with only one center ($M^{\rm eff}=M$), its position
has not to be coincident with the center of original one,
$\Delta\eta$. In this case, $M^{\rm eff}\ne M$.
\end{enumerate}

Evidently, these facts are able to explain the exponential
increasing of FM ({see} {Fig.~\ref{fig12}} later). The effects of
$M^{\rm eff}$ cannot be neglected in such large interacting
system like Pb+Pb, since the degree of upwards increasing
strongly depends on the number of participating projectile
nucleons and on the size of the target tubes \cite{TawP13}.

\subsection{Influence of $\gamma$ conversion on 1D FM}
\label{sec:gamma1d}

The measured FM are expected to be partially influenced by the
$e^-$-$e^+-$pairs produced by Dalitz decay and/or by
$\gamma$-conversion. In \cite{TawDs}, the probabilities of
$\gamma$-conversion in the emulsion and in the lead foil are
calculated and found that the percentage of $\gamma$'s converting
to electron-pairs in the lead foil is $\sim 1.71\pm0.1$ and in
the emulsion is $\sim 0.48\pm 0.06$. Taking the detector
resolution and the possible space angles between such pairs into
account, it is possible to estimate that $\geq 78\%$ of these
electrons cannot be grabbed by MIRACLE Lab. In addition to this
effect of {\it unresolved close pairs}, most of, if not all, the
rest of these electrons are obviously included within the
so-called $4\%$ discrepancy (Section \ref{sec:2}). Therefore, we
could consider that there is a negligible effect of $\gamma$'s on
1D FM. Another argument to support this conclusion is coming out
from the simulation processes. Through adding $\gamma$'s to
FRITIOF for example, it was found \cite{emu0192}: First, the
influence from the Dalitz decay is obviously negligible. Second,
the influence from the $\gamma$-conversion is undersized, or at
least, comparable with the quoted statistical acceptances.
Therefore, we can conclude that both of Dalitz decay and
$\gamma$-conversion have an infinitesimal effect on 1D FM,
especially in the data retrieved by MIRACLE Lab.

\section{Results in Multi-Dimensions}
\label{sec:5}

\begin{figure}[htb]
\vspace*{-6pt}
\begin{minipage}[t]{55mm}
{\quad\ \epsfysize=7cm\epsfbox{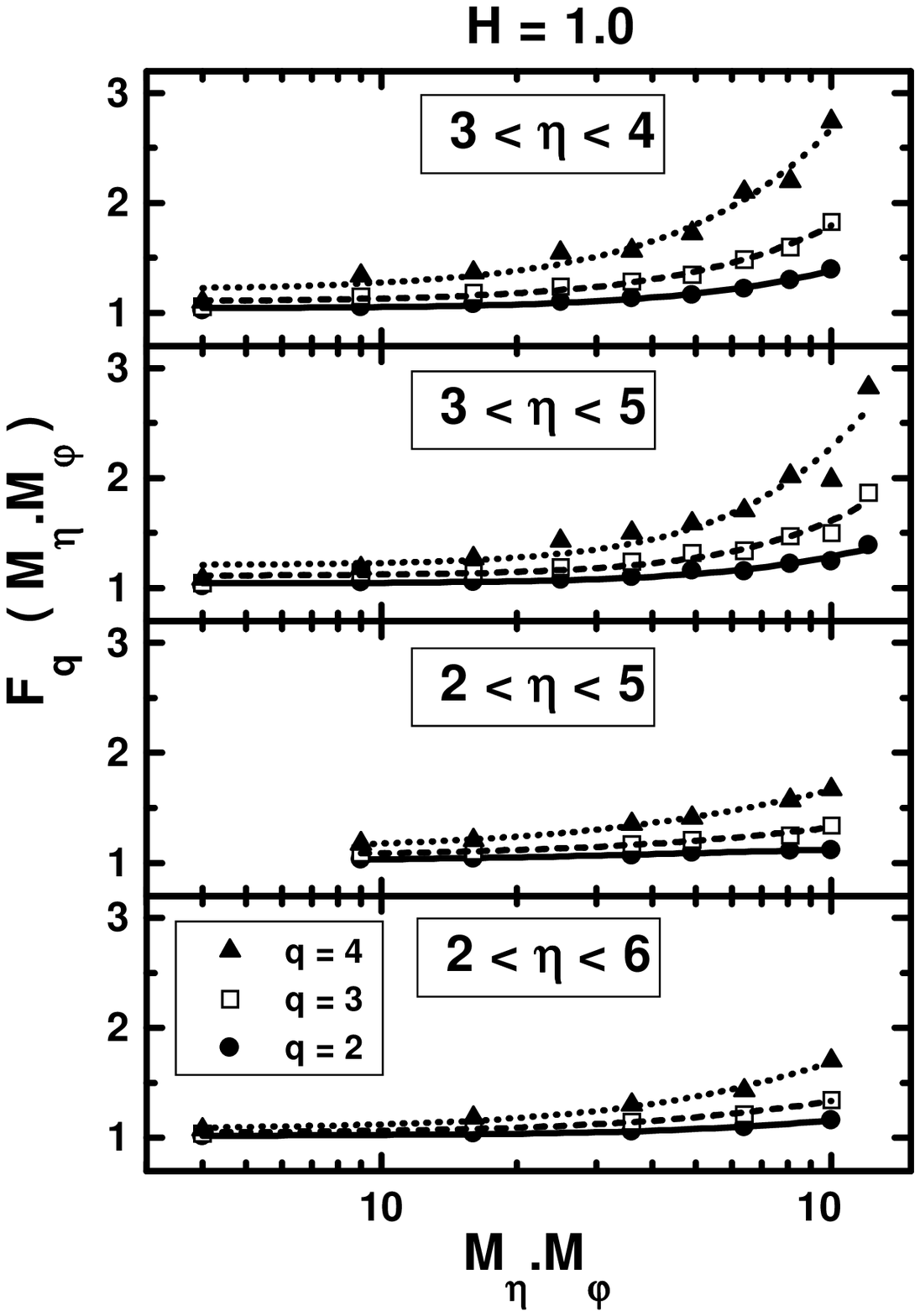}}
\caption[]{\small \it FM is performed here in two-dimensions
  simultaneously and then compared in the different
  $\Delta\eta$-intervals. The partition process is performed for {\it small} number
  $M_{\eta}\cdot M_{\phi}$. Here, FM apparently increase faster than the 1D FM
  ({Fig.~\ref{fig1}} and {Fig.~\ref{fig2}}) with the increasing of
  $M_{\eta}\cdot M_{\phi}$.
\label{fig8}}
\end{minipage} \hfill
\vspace*{-104mm} \hspace*{6cm} \hfill
\begin{minipage}[t]{55mm}
{\quad\ \epsfysize=65mm\epsfbox{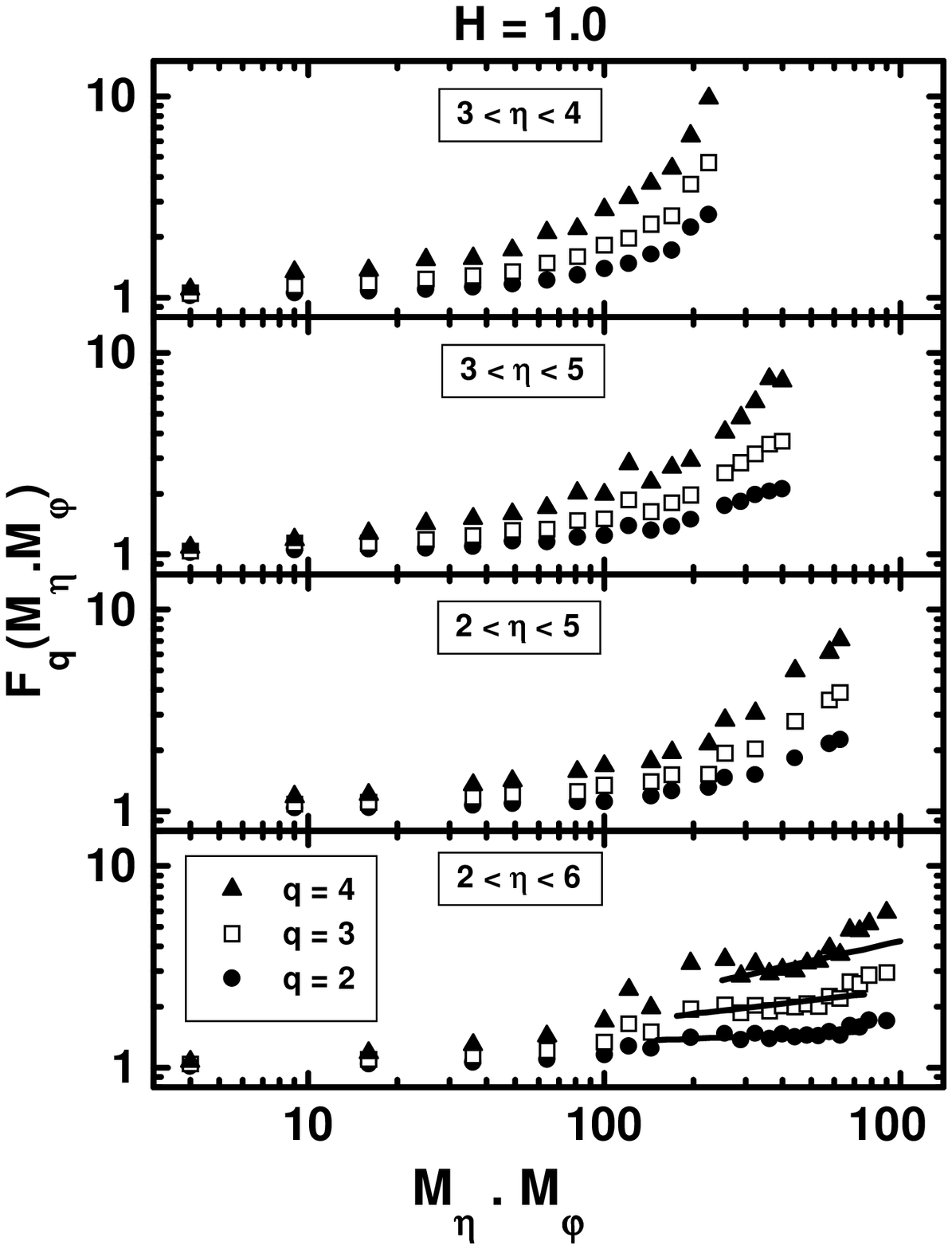}}
\vspace*{8pt}
\caption[]{\small \it The same
  as in {Fig.~\ref{fig8}}, but for {\it large} $M_{\eta}\cdot
  M_{\phi}$. We notice that the increasing of \hbox{$\log F_q$ vs.
  $\log (M_{\eta}\cdot M_{\phi})$} is upwards binding. Saturation
  regions as in {Fig.~\ref{fig3}} are observed here, especially in the
  largest $\Delta\eta$-interval.
\label{fig9} }
\end{minipage}
\vspace*{24pt}
\end{figure}

For clear understanding of the dynamics of the interaction and to
investigate the different reasons responsible for the multiplicity
fluctuations and consequently to explain the power-law behavior at
the final state of particle production, it is necessary to
investigate FM in multi-dimensions ($\phi$, $p_t$, $\eta$)
\cite{Ochs90}. Because of the absence of any external magnetic
field during the emulsion exposure (Section \ref{sec:2}), it is
not possible to measure $p_t$. It remains only the two dimensions,
$\eta$ and $\phi$ \cite{emu0192}. For the multi-dimensional study
of FM, we should first define the so-called {\it
multi-dimensional} partition method. In next subsection, we
introduce methods generally used for the two-dimensional
partition.

\subsection{Methods for two-dimensional partition}
\label{sec:51}

If the $\Delta\eta$-interval is divided into the same number of
bins as $\Delta\phi$, this method is called {\it self-similar}
partition, i.e.\ the partition in certain dimension controls the
method of the partition in the other one. Clearly, this method
leads to total partition number of $M^2$. A practical method to
contemporaneously determine the partitions in two-dimensions is
given by the so-called {\it Hurst-exponents} \cite{Mand91}, \be
{\cal H}_{12} = \frac{\ln M_1}{\ln M_2}. \label{e:12} \ee $M_i;
\;  i=\{1,2\}$ is the number of bins in $i$-th direction. If
${\cal H}=1$, the two variables, $\eta$ and $\phi$, will be
divided in a {\it self-similar} way. It was pointed out in
\cite{emu012,Hov69} and discussed in Section \ref{sec:41} that
the superposition in $\eta$-dimension makes the {\it effective}
partition number, $M^{\rm eff}_{\eta} \gg M_{\eta}$. In the
contrast, the superposition in $\phi$-dimension makes no change
at all, i.e.\ $M^{\rm eff}_{\phi} = M_{\phi}$. These lead to
${\cal H}_{\eta\phi} = \ln M_{\eta}^{\rm eff}/\ln M_{\phi}$, and

\be M_{\eta}^{\rm eff} = e^{{\cal H}_{\phi\eta} \ln M_{\phi} }.
\label{e:13} \ee

To study the possible {\it abnormal} behavior of FM  with $M$, we
should perform the analysis of FM for only one value of ${\cal
H}$. Otherwise, the {\it observed} trend of the calculated FM
will always be bending upwards \cite{Yuan93}, even if there is no
fluctuation in the particle production. The upwards increasing of
FM which is depending only on the partition method, can be
weakened or totally removed ({review} {Eq.\ (\ref{e:13})}), if
the exponents, ${\cal H}$, are given a suitable value. At this
position, it is worthwhile to refer to the conclusion of
\cite{TawP13} that the exponents, ${\cal H}$, applied in the
particle production strongly depend on the energy and the mass of
projectile and target.

\subsection{Results in two-dimensions}
\label{sec:52}

$F_q$ are drawn in {Fig.~\ref{fig8}} as functions of
$M_{\phi}M_{\eta}$ for the different $\Delta\eta$ intervals. The
partition processes have been performed in a {\it self-similar}
way. The number of partition, $M_{\phi}M_{\eta}$, goes up to
$100$ for the different $\Delta\eta$ intervals, which enable us
to study FM for different bin sizes. Here there is no restriction
on the azimuthal angles ($\Delta\phi=2\pi$ and
$\delta\phi\in\{0,2\pi\}$). In addition to the features
concerning with the 2D-analysis, almost similar features as in
{Fig.~\ref{fig1}} and {Fig.~\ref{fig2}} are observed here, for
example, the reduction of FM with the increasing of the
considered $\Delta\eta$-window and the large dispersions of FM
with the decreasing of $\Delta\eta$. Obviously, the exponential
increase of FM ({Fig.~\ref{fig9}}) can be proportionally
connected with $\Delta\eta$-widths ({see} {Fig.~\ref{fig3}} and
{Fig.~\ref{fig4}}). The last could be understood, according to
the facts that the increase of $\Delta\eta$-width leads to an
increasing of the size of the resulted bin $\delta\eta$. These
observations support the assumption that the upwards increasing
is to be explained due to the superposition of the non-coincident
sub-intervals of the {\it sub} pseudo-rapidity values
\cite{emu012,Yuan93} (Section \ref{sec:41}). The partition
numbers drawn here are smaller than the effective one, $M^{\rm
eff}$. According to {Eq.\ (\ref{e:13})}, $M^{\rm eff}$ are
functions of the Hurst-exponents ${\cal H}$ (Section
\ref{sec:51}). Continue the partition process in
{Fig.~\ref{fig9}}, FM keep on the upwards increasing. As noticed
in {Fig.~\ref{fig3}}, saturation regions get formed in wide
$\Delta\eta$-intervals. In bottom picture of {Fig.~\ref{fig9}},
the saturation curves are completely formed.

\subsection{Influences of $\gamma$ conversion on 2D FM}
\label{sec:gamma2d}

As discussed in Section \ref{sec:gamma1d}, most of
$\gamma$-conversions are expected to take place within the lead
foil. The measuring system MIRACLE Lab is able to percolate the
measurements from most of the tracks of electron-pairs (Section
\ref{sec:2} and~\ref{sec:gamma1d}). Generally, the contributions
of $\gamma$'s to multi-dimensional FM \cite{emu0192,Albr94} have
to be taken into account. The features of MIRACLE Lab discussed
above, for which we concluded that the effects of $\gamma$'s on
1D FM are to be neglected, are still valid also in 2D FM. Since we
noticed that the interacting system, Pb+Pb, is intermittent in 1D
FM ({Fig.~\ref{fig1}} to {Fig.~\ref{fig4}} and {Fig.~\ref{fig8}}
to {Fig.~\ref{fig9}}), we do not constrain to go through the
estimation of $\gamma$'s in 2D \cite{Ochs912} again. Another
reason for this renounce is the fact that the estimation of
$\gamma$'s is traditionally performed in the following way:
Adding these processes to the used simulating code and comparing
the experimental results with the simulated ones. If the last are
able to describe the first, the fluctuations of $\gamma$'s are
the responsible source for power-law behavior of FM
\cite{emu0192,Albr94}. In our case, It is not needed to process
this method, since the used measuring system is obviously able to
filter the measurements from the electron-pairs.

\begin{figure}[h]
\vspace*{-6pt}
\hspace*{-6pt}
\begin{center}
\begin{tabular}{cc}
\epsfxsize=6.06cm\epsffile{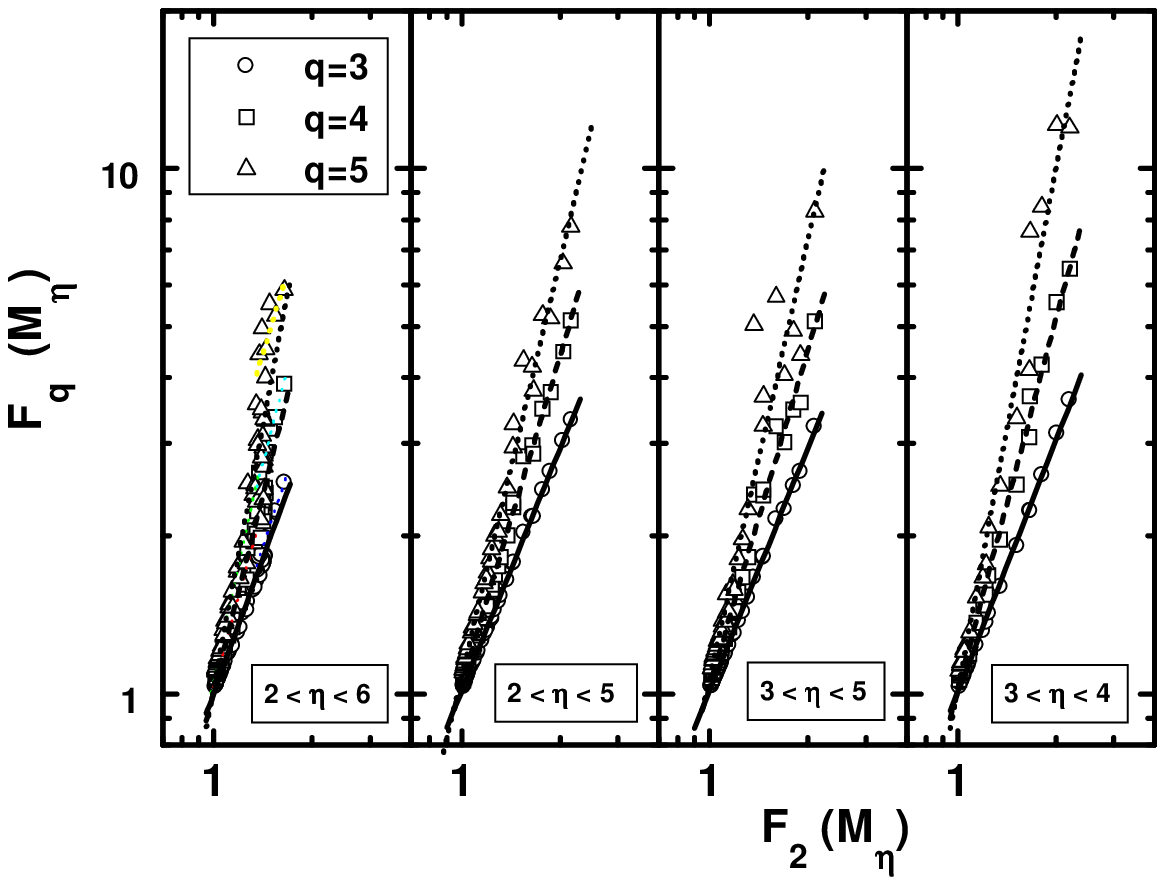}\hspace*{-16pt}  &
\epsfxsize=6.46cm\epsffile{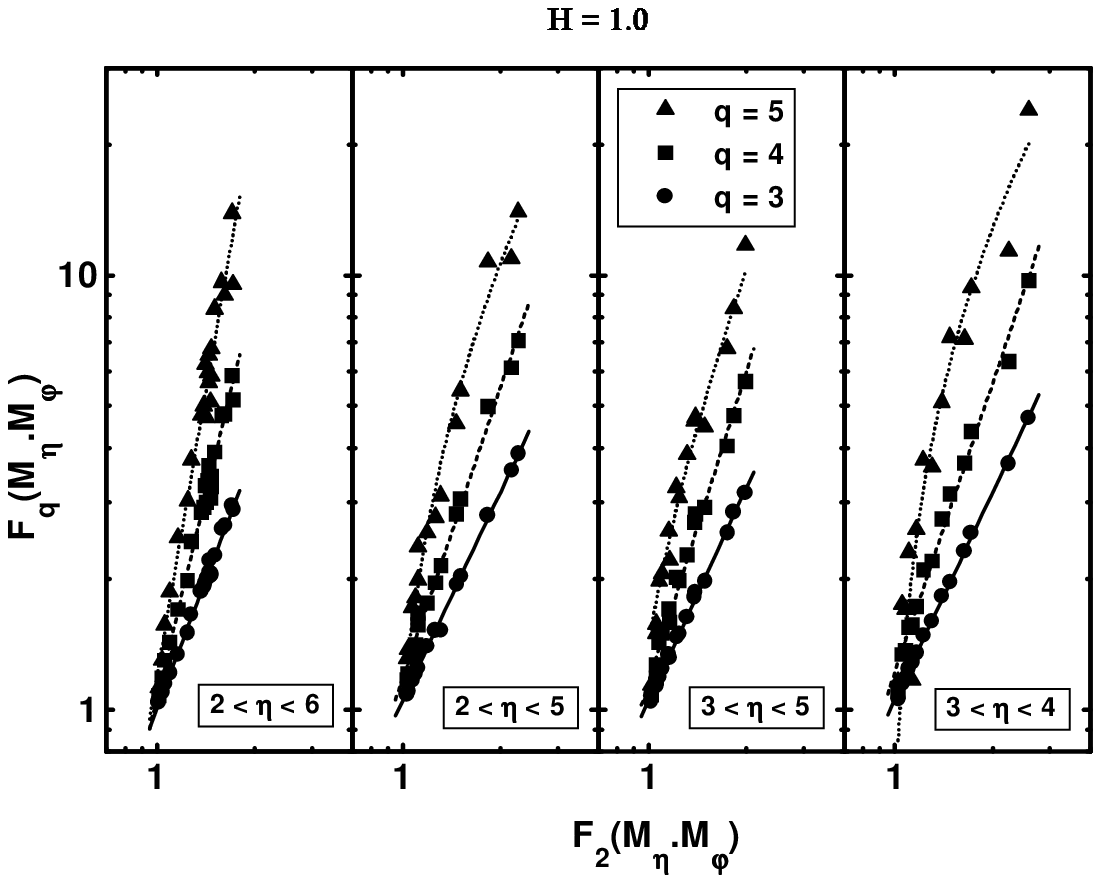} \\
\end{tabular}
\vspace{-16pt}
\end{center}
\caption[]{\small \it The dependence of $q$- on second-order FM in
$\eta$-dimension is depicted. The lines represent the linear
fitting of the experimental data. In {Table~\ref{t:1}}, the
fitting parameters are listed. We notice a positive dependence of
$F_q$ on $F_2$ for all $\Delta\eta$-intervals in one- as well as
in two-dimensions. \label{fig1011}}
\vspace*{20pt}
\end{figure}

\section{L\'evy Indices and Anomalous Dimensions}
\label{sec:levy}

\begin{figure}[thb]
\vspace*{6pt}
\epsfxsize=11.5cm
\begin{center}
\epsfbox{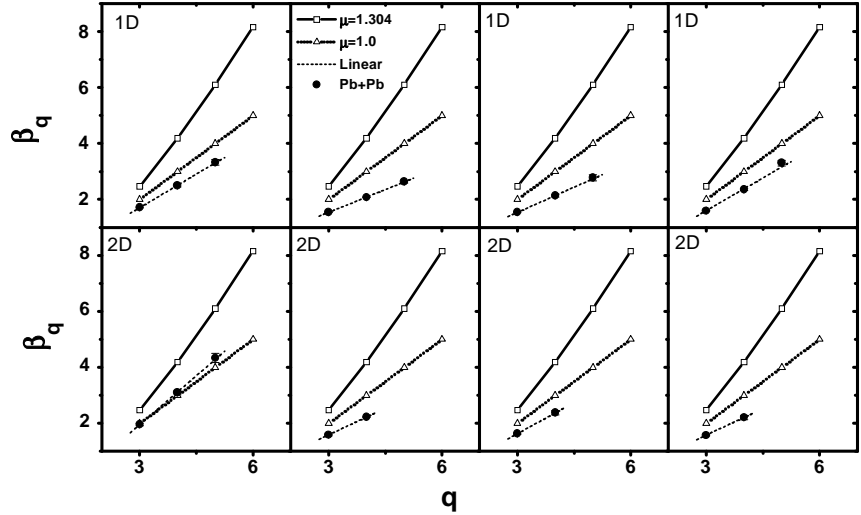} \end{center} \vspace*{-5mm} \caption[]{\small \it The
ratios of intermittence exponents, $\beta_q=\phi_q/\phi_2$, given
as functions of orders, $q$. The solid lines represent the fitting
according to the relation (\ref{e:19}). The fitting according to
relation (\ref{e:17}) is represented by the dotted lines. The
conventional linear power fitting are given by the thin
dash-lines. The experimental data are represented by the solid
points. \label{fig12}}
\vspace*{20pt}
\end{figure}

\subsection{Ratios of intermittence exponents $\phi_q/\phi_2$}
\label{sec:ratios}

\begin{table}
\vspace*{-12pt}
{\baselineskip=11pt \caption[]{\small \it The fit parameters,
$\phi_q/\phi_2$ and $c_q$, for the straight lines drawn in
{Fig.~\ref{fig1011}}, left picture for 1D analysis, are listed
here. \baselineskip=10pt}\label{t:1} }
\vspace*{-0.5cm}
\begin{center}
\begin{tabular}{cccc  cccc} \hline\\[-10pt]
$\eta$ & q &  & value &   $\eta$ & q &  & value\\[2pt]
\hline\\[-10pt]
           & & $c_3$ & $0.003\pm0.001$ &  &   & $c_3$ & $0.005\pm0.001$ \\
           &\raisebox{1.4ex}[-1.6ex]{3} & $\phi_3$/$\phi_2$ & $1.598\pm0.009$ &
       &\raisebox{1.4ex}[-1.6ex]{3} & $\phi_3$/$\phi_2$ & $1.543\pm0.007$ \\
           &   & $c_4$  & $0.006\pm0.004$ &
           &   & $c_4$  & $0.015\pm0.003$ \\
\raisebox{1.4ex}[-1.6ex]{$3<\eta<4$} &\raisebox{1.4ex}[-1.6ex]{4}
&
           $\phi_4$/$\phi_2$ & $2.354\pm0.033$ &
\raisebox{1.4ex}[-1.6ex]{$3<\eta<5$} &\raisebox{1.4ex}[-1.6ex]{4}
&
           $\phi_4$/$\phi_2$ & $2.131\pm0.035$\\
           &   & $c_5$  & $0.007\pm0.014$  &  &  & $c_5$  & $0.031\pm0.011$\\
           &\raisebox{1.4ex}[-1.6ex]{5} & $\phi_5$/$\phi_2$ & $3.300\pm0.102$ &
       &\raisebox{1.4ex}[-1.6ex]{5} & $\phi_5$/$\phi_2$ & $2.777\pm0.110$\\ \hline
           &   & $c_3$  & $0.006\pm0.001$ &  &  & $c_3$ & $0.001\mp0.001$ \\
           &\raisebox{1.4ex}[-1.6ex]{3} & $\phi_3$/$\phi_2$ & $1.537\pm0.004$ &
       &\raisebox{1.4ex}[-1.6ex]{3} & $\phi_3$/$\phi_2$ & $1.709\pm0.023$ \\
           &   & $c_4$  & $0.017\pm0.002$ &
           &   & $c_4$  & $0.004\pm0.005$ \\
\raisebox{1.4ex}[-1.6ex]{$2<\eta<5$} &\raisebox{1.4ex}[-1.6ex]{4}
&
           $\phi_4$/$\phi_2$ & $2.083\pm0.015$ &
\raisebox{1.4ex}[-1.6ex]{$2<\eta<6$} &\raisebox{1.4ex}[-1.6ex]{4}
&
           $\phi_4$/$\phi_2$ & $2.495\pm0.056$ \\
           &   & $c_5$  & $0.035\pm0.004$ & & & $c_5$  & $0.011\pm0.013$ \\
           &\raisebox{1.4ex}[-1.6ex]{5} & $\phi_5$/$\phi_2$ & $2.631\pm0.039$ &
       &\raisebox{1.4ex}[-1.6ex]{5} & $\phi_5$/$\phi_2$ & $3.321\pm0.129$
           \\ \hline
\end{tabular}
\end{center}
\end{table}

The relation between $q$- and second-order FM is given as: \be
\log F_q(M_{\{\eta|\phi\}}) = \frac{\phi_q}{\phi_2} \log
F_2(M_{\{\eta|\phi\}}) + c_q, \label{e:14} \ee $\phi_2$ and
$\phi_q$ are the slopes of the relations between $\log
F_2(M_{\{\eta|\phi\}})$ and $\log F_q(M_{\{\eta|\phi\}})$ and the
corresponding $M_{\{\eta|\phi\}}$. $c_q$ are constant. From these
two figures (also review {Fig.~\ref{fig1}} and {Fig.~\ref{fig2}}
again), we notice that the exponents, $\phi_q$, do not depend on
$M_{\{\eta|\phi\}}$. Therefore, in practice the slope ratios,
$\phi_q/\phi_2$, can be deduced from the relation $\log
F_q(M_{\{\eta|\phi\}})$ vs. $\log F_2(M_{\{\eta|\phi\}})$.
{Fig.~\ref{fig1011}} depicts such relations for the orders,
$q=\{3,4,5\}$. The lines represent the power fitting of the
experimental data. Orders $q$ increase, the slopes and their
corresponding statistical errors get large. Comparisons between
the fitting parameters in the different $\Delta\eta$-intervals are
summarized in {Table~\ref{t:1}}. The results from the 2D-analysis
are graphically illustrated in the right part of
{Fig.~\ref{fig1011}} and their corresponding parameters are
listed in {Table~\ref{t:2}}. From these two tables and by taking
into account the different statistical errors in 1D and 2D, we can
conclude that the corresponding slope ratios $\phi_q/\phi_2$
smoothly get larger for higher analysis dimension. In left
picture in {Fig.~\ref{fig1011}}, {\it especially} in the largest
$\Delta\eta$-interval, $2<\eta<6$, we notice the existence of two
separated regions. Each of them has been fitted and depicted as
thin line. The suddenly increasing of the ratios, $\phi_q/\phi_2$,
at the boarder between these two regions measures the maximum
amplitude of the saturation plateau of second-order FM ($\sim
1.5$). At this point, $\phi_q/\phi_2\rightarrow\infty$.  The
comparison between one- and two-dimensional FM requires to give
${\cal H}$-exponents, used to define the method of partition in
two-dimensions, a suitable constant value. The value, ${\cal
H}=1$, obviously cannot overcome the effects of the superposition
of the sub-intervals ({\it elementary collisions}). Therefore, we
expect that FM exponentially increases for $\delta\eta\rightarrow
0$ ({Fig.~\ref{fig9}}). This clearly explains the non-linear
behavior of parts of the relations shown in the right picture in
{Fig.~\ref{fig1011}}.

\begin{table}
\vspace*{-12pt} {\baselineskip=11pt \caption[]{\small \it The parameters,
$\phi_q/\phi_2$ and $c_q$, from the linear power fitting drawn in
{Fig. ~\ref{fig1011}}, right picture for the 2D analysis, are
given here.\baselineskip=10pt}\label{t:2}}
\vspace*{-0.5cm}
\begin{center}
\begin{tabular}{cccc cccc} \hline\\[-10pt]
$\eta$ & q &  & value &   $\eta$ & q &  & value\\[2pt]
\hline\\[-10pt]
           & & $c_3$ & $0.024\pm0.003$ &  &   & $c_3$ & $0.021\pm0.003$ \\
           &\raisebox{1.4ex}[-1.6ex]{3} & $\phi_3$/$\phi_2$ & $1.579\pm0.017$ &
       &\raisebox{1.4ex}[-1.6ex]{3} & $\phi_3$/$\phi_2$ & $1.629\pm0.023$ \\
           &   & $c_4$  & $0.084\pm0.014$ &
           &   & $c_4$  & $0.064\pm0.012$ \\
\raisebox{1.4ex}[-1.6ex]{$3<\eta<4$} &\raisebox{1.4ex}[-1.6ex]{4}
&
           $\phi_4$/$\phi_2$ & $2.213\pm0.077$ &
\raisebox{1.4ex}[-1.6ex]{$3<\eta<5$} &\raisebox{1.4ex}[-1.6ex]{4}
&
           $\phi_4$/$\phi_2$ & $2.377\pm0.083$\\
           &   & $c_5$  &   &  &  & $c_5$  &  \\
           &\raisebox{1.4ex}[-1.6ex]{5} & $\phi_5$/$\phi_2$ &  &
       &\raisebox{1.4ex}[-1.6ex]{5} & $\phi_5$/$\phi_2$ & \\ \hline
           &   & $c_3$  & $0.021\pm0.005$ &  &  & $c_3$ & $0.002\mp0.006$ \\
           &\raisebox{1.4ex}[-1.6ex]{3} & $\phi_3$/$\phi_2$ & $1.592\pm0.028$ &
       &\raisebox{1.4ex}[-1.6ex]{3} & $\phi_3$/$\phi_2$ & $1.965\pm0.040$ \\
           &   & $c_4$  & $0.072\pm0.010$ &
           &   & $c_4$  & $0.024\pm0.012$ \\
\raisebox{1.4ex}[-1.6ex]{$2<\eta<5$} & \raisebox{1.4ex}[-1.6ex]{4}
&
           $\phi_4$/$\phi_2$ & $2.226\pm0.060$ &
\raisebox{1.4ex}[-1.6ex]{$2<\eta<6$} & \raisebox{1.4ex}[-1.6ex]{4}
&
           $\phi_4$/$\phi_2$ &  $3.102\pm0.081$ \\
           &   & $c_5$  &    &  & & $c_5$ & $0.075\pm0.025$ \\
           &\raisebox{1.4ex}[-1.6ex]{5}   & $\phi_5$/$\phi_2$ &  &
           &\raisebox{1.4ex}[-1.6ex]{5}   & $\phi_5$/$\phi_2$ & $4.337
                                             \pm0.170$ \\ \hline
\end{tabular}
\end{center}
\end{table}

\subsection{Physics of power-law behavior}
\label{sec:FM-sour}

As mentioned in Section \ref{sec:1} and discussed elsewhere
\cite{Bial19,Brax91,Ochs911}, the sources responsible for the
power-law behavior at the final state of particle production can
be classified into two main groups:
\begin{enumerate}
\item {\it Self-similar processes.} By using Gaussian distributions
      \cite{Bial19}, the ratios of $q$- to second-order intermittence exponents
      can be expressed as:

\be
\frac{\phi_q}{\phi_2}=\frac{d_q}{d_2}(q-1)=\left(\begin{array}{c}
             q \\ \\ 2 \end{array}\right).
\label{e:15}
\ee

      As discussed in \cite{Brax91,Ochs911}, the implementation of L\'evy
      index, $\mu$ \cite{Gned60}, leads to:

\be \frac{d_q}{d_2} =
\frac{q^{\mu}-q}{2^{\mu}-2}\cdot\frac{1}{q-1}, \label{e:16}
\ee

      where $\mu$ has a continuous spectrum, $ 0 <
      \mu\le 2$, ({\it region of L\'evy stability}). Experimentally, EHS/NA22
      collaboration \cite{Agab93} found that the index $\mu$, in
      principle exceeds the upper boundary, 2.
      As we will see later, L\'evy index within its
      restricted stability region evidently is not able to describe the
      experimental relations between $d_q/d_2$ and the orders $q$.
      Only for 2D FM in the rapidity region ranging
      from 2 to 6, we get a stable value, $\mu=+0.1$
({Fig.~\ref{fig13}}). \\
Relation (\ref{e:15}) is evidently an approximation of anomalous
ratios, $d_q/d_2$. Obviously, this relation is not satisfactorily
applicable in the distribution tails. Meanwhile relation
(\ref{e:16}) can effectively determine these tails. As given
above, L\'evy index has two boundaries (drawn in
{Fig.~\ref{fig13}} as dotted lines):

\begin{figure}[htb]
\vspace*{-6pt}
\epsfxsize=11.5cm
\begin{center}\epsfbox{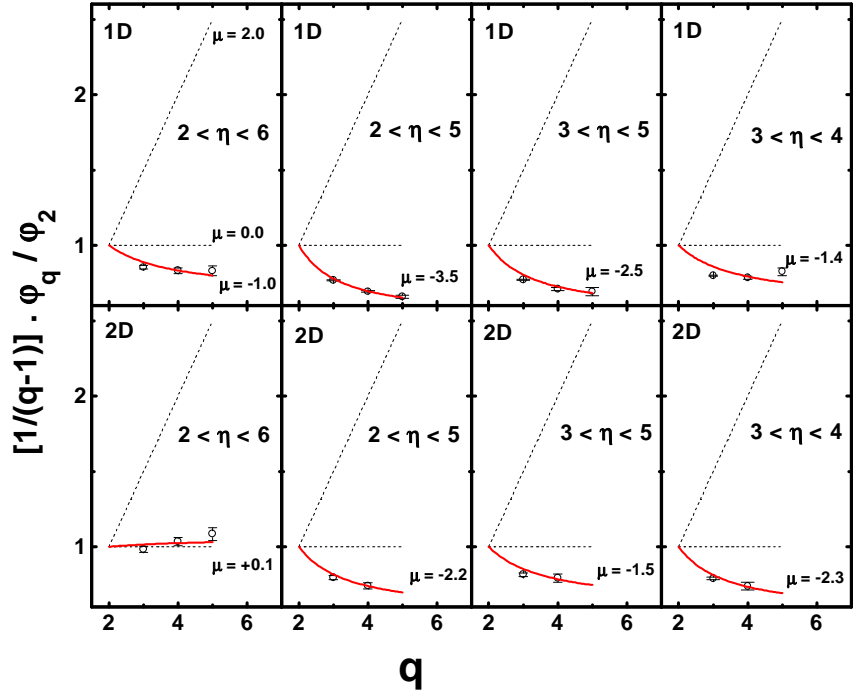}\end{center}
\vspace*{-5mm}
\caption[]{\small \it Dependencies of the anomalous
  dimension ratios, $\phi_q/\phi_2$, with the orders, $q$, for the
  different $\Delta\eta$-intervals and for the possible two dimension,
  $\eta$ and $\phi$. The dotted lines represent the two boundaries of
  L{\'e}vy stable region. The experimental results are represented by
  the open circles. The Solid lines are the solutions of {Eq.\
    (\ref{e:16})} for the given $\mu$-values. 
\label{fig13} }
\vspace*{20pt}
\end{figure}

\begin{figure}
\vspace*{-6pt}
\epsfysize=8cm
\begin{center}\epsfbox{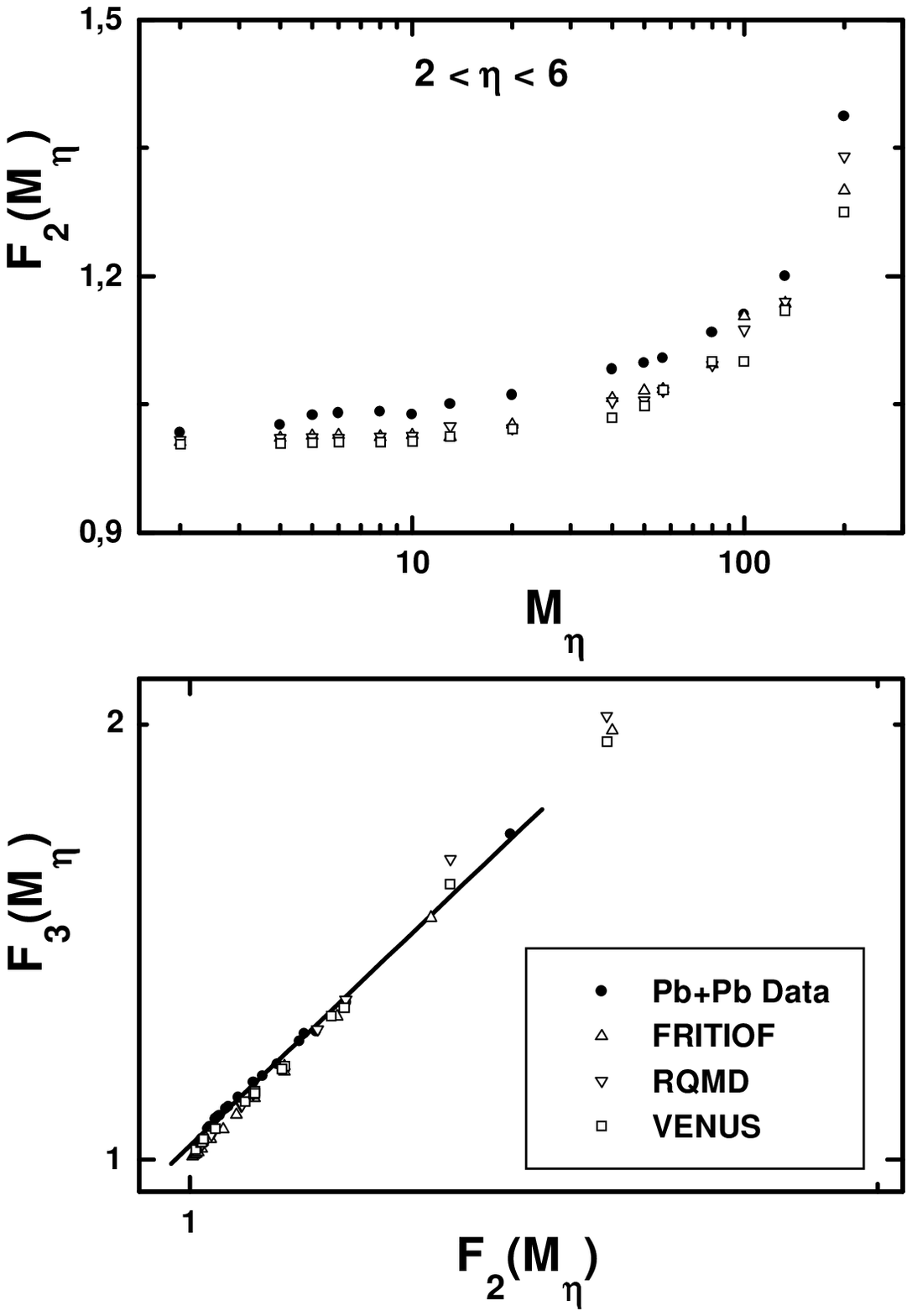}
\end{center}
\vspace*{-0.5cm}
\caption[]{\small \it Comparison between the
  second-order FM and the partition number, $M_{\eta}$, for the
  experimental and simulated events (Section \ref{sec:21}). The
  disagreement between the data and the three models becomes larger
  for larger $M_{\eta}$. In bottom part, the third-order FM are given
  as functions of the second-order ones for the same data set as in
  the top picture. The line represents the linear power fitting of the
  experimental data. Generally, we conclude that non of the three
  models is able to describe the experimental results.
\label{fig14} }
\vspace*{20pt}
\end{figure}

\begin{enumerate}
\item $\mu=2$ leads to $d_q/d_2 = q/2$ corresponding to
      {\it self-similar branching processes} represented by
      the well-known Gaussian distributions, and
\item $\mu=0$ leads $d_q/d_2 = 1$ corresponding to {\it mono-fractal
      behavior} resulted from second-order phase-transition
      from QGP to hadron-gas \cite{Heg93}.
\end{enumerate}
Between these two limits, {Eq.\ (\ref{e:15})} is not valid and  should
be replaced by the relation (\ref{e:16}).
\item If QGP indeed is to be produced in the relativistic
  heavy-ion collisions, the interacting system has to suffer
  {\it second-order phase-transition} during its space-time evolution.
  In this case the observed intermittence behavior
  (in form of power-scaling behavior)
  can be compared with the behavior in a system of mono-fractal
  ({see} {Fig.~\ref{fig12}} and {Fig.~\ref{fig13}}),

\be \frac{\phi_q}{\phi_2} = q-1. \label{e:17} \ee
\end{enumerate}

Then we can conclude that the dependence of the ratios,
$\phi_q/\phi_2$, on the orders, $q$, reflects essential
information about the reaction dynamics \cite{Bial19}. In
following sections, we will study the different intermittence
ratios in relation with the orders, $q$.

\subsection{Relations between $\phi_q/\phi_2$ and the orders $q$}
\label{sec:61}

In form of a power-law, relation (\ref{e:14}) can be re-written
as:

\be F_q \propto F_2^{\beta_q}; \;\;\;\;\;\;
\beta_q\equiv\phi_q/\phi_2\,. \label{e:18} \ee

It is clear to recognize that the critical powers, $\beta_q$, are
independent on the partition process or on the analysis
dimension. This power-law is valid in one- as well as in
multi-dimensions \cite{Ochs911}. In \cite{Hwa92}, the predicted
second-order phase-transition in heavy-ion collisions has been
simulated by using a specific form of Ginzburg--Landau model
\cite{GL80}. It has been found that critical exponents,
$\beta_q$, are independent on the temperature, $T$, just below the
critical one, $T_c$. Therefore, the behavior of $\beta_q$
effectively characterizes the phase-transition to coherent light
and also its aftereffects

\be \beta_q = (q-1)^{\nu}; \;\;\;\;\;\;  \nu=1.304 \,.
\label{e:19} \ee

In our case, the aftereffects is the ratios, $\phi_q/\phi_2$, or
fundamentally the fluctuations in the particle production. $\nu$
are not looking like the conventional critical indices describing
only the behavior of quantities near the critical point. In
addition to this, they are independent on the details of the
interacting system and therefore can effectively characterize the
behavior of all measurable quantities \cite{Hwa92}
($\phi_q/\phi_2$, etc).

{Figure \ref{fig12}} describes the powers, $\beta_q$, as functions
of the orders, $q$. As given before, $\beta_q$ can directly be
determined from the slopes of straight-line portions in
{Fig.~\ref{fig1011}}. The slopes are given in {Tables~\ref{t:1}}
and {\ref{t:2}} for 1D and 2D, respectively. The lines with empty
squares are the fitting according to the relation (\ref{e:19}).
Almost the same process, but for $\nu=1.0$, is represented by the
dash-lines. The value $\nu=1.0$ in {Eq.\ (\ref{e:17})} is
originally obtained from the intermittent behavior of 2D
Ising-model \cite{Satz89} and also from the thermal second-order
phase-transition (Section \ref{sec:ratios}). From
{Fig.~\ref{fig12}}, we also notice that the first fitting is
unable to describe the experimental data. Although the second one
results lines close to the experimental points, it remains unable
to completely predict the experimental results. Only, 2D-analysis
in the region, $2<\eta<6$, produces one line coincident with the
points representing the experimental data (solid points).

\subsection{Relations between $d_q/d_2$ and the orders $q$}
\label{sec:62}

{Figure \ref{fig13}} depicts the ratios, $d_q/d_2$, in dependence
with the orders, $q$, for the different $\Delta\eta$-intervals.
The relations between the anomalous and R{\'e}nyi dimensions are
given in Section \ref{sec:31}. Virtually, this figure can be used
to distinguish between the two possible sources of the
intermittent behavior discussed in Section \ref{sec:FM-sour}. As
we can see, only L\'evy indices ranging from $-1.0$ to $-3.5$ can
fit our experimental data. The most acceptable value, $\mu=+0.1$,
is obtained for 2D FM in the largest $\Delta\eta$-interval. The
reasons and interpretations for this mono-fractal behavior
($\mu\rightarrow 0$, {Eq.\ (\ref{e:16})}) we will discuss later.
Again, the predictions of {Eq.\ (\ref{e:17})} are exactly the
lower L\'evy boundary, where the line $\mu=0$ is.

\section{Remarks and Final Conclusions}
\label{sec:7}

The following remarks have to be considered for the {\it evidence}
of the thermal phase-transition noticed in this article:
\begin{enumerate}
\item FM have to be re-performed for higher orders and dimensions.
      Such a way, we get more lines for {Fig.~\ref{fig12}} and
{Fig.~\ref{fig13}},
      which might support the observation of thermal phase-transition.
      For the same reason, more measurements are also required.
\item 2D FM have to be re-investigated by using other values of
      ${\cal H}$. Such a way, we can investigate the scaling-law
      behavior of FM away from the {\it non-canonical} effects stemming only from
      the partition method itself (Section \ref{sec:51}).
\item $\Delta\eta$-windows have to re-determined to specify the
      regions which fulfill the requirement of L\'evy stability. The
      non-considered interval might include {\it leading baryons} with
   pseudo-rapidity values close to that of the projectile nucleons. In that case,
   most of particles are of course fragmentations. Therefore, they behave
   differently than the produced ones ($\pi$'s).
\item In forthcoming steps, more weight should be given to the
      smaller intervals and to the dynamics of the interaction.
      The last could be studied by using the {\it event-by-event}
      fluctuations. Generally, the analysis of single events with huge
   multiplicity enable us to study different physics: geometry of
   interaction, space-time evolution, two-particle
      correlation functions, microscopic and  thermodynamic
      collisions, QCD fluctuations, phase-transition, etc. \cite{NA49}.
\end{enumerate}

Even, if we could include {\it leading}-particles, it is still
early to dwell on ultimate conclusion. Since these {\it
non-produced} particles may uncommonly comport as the produced
ones, $\pi$'s. Although this conclusion could, in the one hand, be
a uproarious one, especially in light of the remarks stated
before, in the other hand, we cannot totally neglect the
demonstrative evidence noticed in {Fig.~\ref{fig13}} and
{Fig.~\ref{fig14}}. Experimentally --- in only one of
$\Delta\eta$ regions --- we have a stable L\'evy index, $\mu=+0.1$
({Fig.~\ref{fig12}} and {Fig.~\ref{fig13}}). This
positive {\it small} value is standing as a signature for the
thermal phase-transition \cite{Brax91}. The other values, $\mu<0$,
indicate that L\'evy indices cannot effectively distinguish
between thermal and non-thermal branching processes.

The stable value, $\mu=+0.1$, is obtained from the two-dimensional
analysis of the largest $\Delta\eta$ region. Within this region,
the effects of the superposition in $\eta$-dimension are much
smaller than that in other $\eta$-intervals \cite{TawP13} (Section
\ref{sec:41}). Within it, there are also {\it non-produced}
particles, which in contrast to the produced particles, are not
sensitive to the phase-transition. As mentioned above, more
measurements, event-by-event analysis and higher dimensions are
highly required to demonstrate such power-law scaling behavior
and then to confirm the evidence of the thermal phase-transition.

\section*{Acknowledgement} This work is financially supported by the
Deutscher Akademischer Austaschdiest ({DAAD}). We are very
grateful to E.\ Stenlund and all colleagues of EMU01 collaboration
for the helpful discussions and the kind assistance. Especially
that they allowed us the use of part of our collaborative
experimental data for this work. A.M.T.\ would like to thank
F.~P{\"u}hlhofer for the continuous support.


\end{document}